# Nanoscale

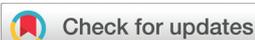

## PAPER

Check for updates

# A classical picture of subnanometer junctions: an atomistic Drude approach to nanoplasmonics†

Tommaso Giovannini, [a] Marta Rosa, [b] Stefano Corni [*,b,c] and Chiara Cappelli [*,a]

The description of optical properties of subnanometer junctions is particularly challenging. Purely classical approaches fail, because the quantum nature of electrons needs to be considered. Here we report on a novel classical fully atomistic approach, ωFQ, based on the Drude model for conduction in metals, classical electrostatics and quantum tunneling. We show that ωFQ is able to reproduce the plasmonic behavior of complex metal subnanometer junctions with quantitative fidelity to full *ab initio* calculations. Besides the practical potentialities of our approach for large scale nanoplasmonic simulations, we show that a classical approach, in which the atomistic discretization of matter is properly accounted for, can accurately describe the nanoplasmonics phenomena dominated by quantum effects.

## 1 Introduction

A cornerstone of nanoscience is that systems at the nanoscale have properties neither of the molecular nor of the macroscopic length scales.[1,2] Nanoplasmonics is a beautiful example of this: localized surface plasmons supported by metal nanostructures disappear in clusters with few atoms, and acquire different properties (surface plasmon polaritons) on extended surfaces.[1–4] The enormous progress of nanoscience has permitted a targeted control of the morphology of nanostructures at the nanometer and even subnanometer scales, thus allowing several applications in plasmonics and nanooptics.[5–9] Most properties of plasmonic nanostructures follow from the tunability of their optical response as a function of their shape and dimensions; in case interparticle gaps are formed, the so-called "hot-spot" regions occur, in which localized surface plasmons can interact with molecules placed in the junctions, allowing single molecule detection.[10–15]

The optical properties of nanostructures are generally treated, independent of the system's size/shape, by resorting to classical approaches.[16–29] However, when the size of the particles or junctions is only a few nanometers or smaller, the quantum nature of electrons emerges, activating quantum tunneling effects across subnanometer interparticle gaps.[19,20,30–32,34–44] Tunnelling effects are not considered in classical models, so that quantum corrected approaches need to be applied.[37,41]

The theoretical study of the atomic-scale features in nanojunctions is still an almost unexplored field, because most phenomenological classical models do not address quantum effects. In fact, as reported by Urbieta *et al.*,[45] an appropriate description of atomic-scale effects would require a full quantum framework, accounting for the atomistic structure of the nanoparticles and the wave nature of electrons building up the plasmonic excitation.

By starting from the above considerations, in this paper we report on a fully atomistic classical model based on three very basic ingredients, *i.e.* the Drude model for conduction in metals, classical electrostatics and quantum tunneling, which is able to reproduce with quantitative fidelity the optical properties of subnanometer junctions. In our approach, which we will call ωFQ (frequency dependent Fluctuating Charges), each atom of the nanostructure is endowed with an electric charge, which is not fixed but can vary as a response to the externally applied oscillating electric field.

Remarkably, here we go a step further with respect to other classical approaches. In fact, we are not using any experimental frequency dependent dielectric constant (possibly corrected for non-locality and electron scattering at the surface), but we allowed the dielectric response of the nanosystem to arise from atom–atom conductivity. Quantum tunnelling effects originate from a geometrical damping imposed on the atom–atom conductivity regime. The model is challenged to accurately reproduce complex *ab initio* simulations on a stretched Na nanorod[46] and two approaching and retracting Na nanoparticles,[47] in which a single atom junction occurs, so that an atomistic description appears to be mandatory. ωFQ is the first ever classical approach succeeding at correctly modeling the optical properties of such systems, which up to now have been successfully treated only at the full DFT level. Remarkably, the results that we will show have relevant practical consequences, because we indeed provide a computationally viable model to investigate subnanometer junctions and complex nanostructures of sizes well beyond what can be currently treated by *ab initio* approaches.

*a Scuola Normale Superiore, Piazza dei Cavalieri 7, 56126 Pisa, Italy.
E-mail: chiara.cappelli@sns.it*
*b Department of Chemical Sciences, University of Padova, via Marzolo 1, Padova, Italy. E-mail: stefano.corni@unipd.it*
*c CNR Institute of Nanoscience, via Campi 213/A, Modena, Italy*

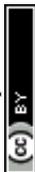

## 2   Theoretical model

The model we are introducing here, ωFQ, has its fundamentals on the Fluctuating Charges (FQ) force field, which is usually adopted for describing molecular systems.[48–53] FQ places on each atom of a molecular system a charge, which is not fixed but allowed to vary as a result of differences in atomic electronegativities. Charges are regulated by atomic chemical hardness, which plays the role of an atomic capacitance. From the mathematical point of view, FQ charges are obtained by minimizing the functional defining the energy of the system (see section S1 given as the ESI†). ωFQ extends the basic formulation of the FQ model to take into account the interaction of the system with an external oscillating electric field $\mathbf{E}(\omega)$. In particular, each atom is assigned a charge, which is allowed to vary as a response of the polarization sources, which also include the external field $\mathbf{E}(\omega)$. Thus, due to the fact that the electric field is a complex quantity, calculated ωFQ charges become complex, being their imaginary value in quadrature with the field (if the field is real) and related to the absorption phenomenon. To build up the ωFQ approach, the time response of charges has to be related to external polarization sources. To this end, two alternative response regimes are set: (i) a conductive regime, in which the exchange of electrons between contiguous atoms is governed by the dynamics of the delocalized conduction electrons, giving rise to a damping; (ii) an alternative conductive regime, in which the exchange of electrons is also mediated by quantum tunneling effects. In this section we briefly discuss the main physical aspects of ωFQ: more details on the derivation of the equations and their implementation are given as the ESI† (see sections S2 and S3).

The first regime is described by reformulating the Drude model of conductance[54] to treat charge redistribution between atoms. The key equation representing the Drude model reads:[54]

$$\frac{d\mathbf{p}}{dt} = \mathbf{E}(t) - \frac{\mathbf{p}}{\tau} \tag{1}$$

where $\mathbf{p}$ is the momentum of the electron and $\tau$ is a friction-like constant due to scattering events. The total charge derivative on atom $i$ can be written as:

$$\frac{dq_i}{dt} = \sum_j A_{ij} \left( n_j \langle \mathbf{p} \rangle \cdot \hat{l}_{ji} - n_i \langle \mathbf{p} \rangle \cdot \hat{l}_{ij} \right) \tag{2}$$

where $A_{ij}$ is an effective area dividing atom $i$ by atom $j$, $n_i$ is the electron density on atom $i$, $\langle \mathbf{p} \rangle$ is the momentum of an electron averaged over the trajectories connecting $i$ and $j$ and $\hat{l}_{ji} = -\hat{l}_{ij}$ is the unit vector of the line connecting $j$ to $i$. By assuming the total charge on each atom to be only marginally changed by an external perturbation, we can assume $n_i = n_j = n_0$. Therefore:

$$\frac{dq_i}{dt} = 2n_0 \sum_j A_{ij} \langle \mathbf{p} \rangle \cdot \hat{l}_{ji} \tag{3}$$

where $\langle \mathbf{p} \rangle \cdot \hat{l}_{ji}$ needs to be estimated. To this end, it is convenient to consider a monochromatic applied electric field, so that eqn (3) translates to:

$$-i\omega q_i = 2n_0 \sum_j A_{ij} \frac{\langle E(\omega) \rangle \cdot \hat{l}_{ji}}{1/\tau - i\omega} \tag{4}$$

To proceed further, $\langle \mathbf{E}(\omega) \rangle \cdot \hat{l}_{ji}$ (the total electric field averaged over the line connecting $j$ to $i$) needs to be connected to atomic properties. This can be done by assuming $\langle \mathbf{E}(\omega) \rangle \cdot \hat{l}_{ji} \approx (\mu_j^{el} - \mu_i^{el})/l_{ij}$, where $\mu_i^{el}$ is the electrochemical potential of atom $i$ and $l_{ij}$ is the distance between atoms $i$ and $j$. Therefore, eqn (4) becomes:

$$\begin{aligned} -i\omega q_i &= \frac{2n_0}{1/\tau - i\omega} \sum_j \frac{A_{ij}}{l_{ij}} (\mu_j^{el} - \mu_i^{el}) \\ &= \frac{2\sigma_0}{1/\tau - i\omega} \frac{A_{ij}}{l_{ij}} (\mu_j^{el} - \mu_i^{el}) \\ &= \sum_j K_{ij}^{dru} (\mu_j^{el} - \mu_i^{el}) \end{aligned} \tag{5}$$

where $n_0 = \sigma_0/\tau$, with $\sigma_0$ being the static conductance of the considered metals. Note that the dependence of both $\sigma_0$ and $\tau$ on the temperature can in principle be considered; however, in this paper the effects of such dependence have not been investigated. Eqn (5) can be rewritten collecting $K_{ij}^{dru} = \frac{2n_0}{1/\tau - i\omega} \frac{A_{ij}}{l_{ij}}$ in a $\mathbb{K}^{dru}$ matrix (see eqn (5)). In order to make the model physically consistent, *i.e.* not to allow electron transfer between atoms that are too far apart, the pairs of atoms considered in eqn (5) have to be selected by exploiting a geometrical criterion, based on $l_{ij}$, *i.e.* to limit the interactions to the nearest neighbors only.

To avoid any issue related to the specific definition of the nearest neighboring atoms, a Fermi-like $f(l_{ij})$ damping function is introduced as a weight of the Drude conductive mechanism:

$$\begin{aligned} -i\omega q_i &= \sum_j (1 - f(l_{ij})) \cdot K_{ij}^{dru} \left( \mu_j^{el} - \mu_i^{el} \right) \\ &= \sum_j K_{ij}^{tot} \left( \mu_j^{el} - \mu_i^{el} \right) \end{aligned} \tag{6}$$

where

$$f(l_{ij}) = \frac{1}{1 + \exp\left[ -d\left( \frac{l_{ij}}{s \cdot l_{ij}^0} - 1 \right) \right]} \tag{7}$$

In eqn (7) $l_{ij}^0$ is the equilibrium distance between the two nearest neighbors in the bulk, whereas $d$ and $s$ are the parameters determining the position of the inflection point and the steepness of the curve. Note that in the case of systems composed of different atomic species, the Fermi-like function defined in eqn (7) needs to be adjusted so to take into account the specificities of the considered material.

Eqn (6) finally defines the ωFQ model. Whenever $f(l_{ij}) = 0$, the purely Drude conductive regime is recovered. For $f(l_{ij}) > 0$, Drude mechanisms exponentially turn off as $l_{ij}$ increases, making electron transfer enter in a second alternative regime. In this regime, the electric current exponentially decreases upon increasing the inter-atomic distance. Therefore, the typical functional form of tunneling exchange is recovered.[37]

Once ωFQ frequency-dependent charges are obtained by solving eqn (6), the complex polarizability $\bar{\alpha}$ is easily calculated. From such a quantity, the absorption cross section is recovered:

$$\sigma_{abs} = \frac{4\pi}{3c}\omega \mathrm{tr}(\mathrm{Im}(\alpha)) \qquad (8)$$

where $c$ is the speed of light, $\omega$ is the external frequency and $\mathrm{Im}(\alpha)$ is the imaginary part of the complex polarizability $\bar{\alpha}$.

The ωFQ approach has been implemented in a standalone Fortran 77 package. Eqn (6) is solved for a set of frequencies given as input. All computed spectra reported in the manuscript were obtained by explicitly solving linear response equations for steps of 0.01 eV. For all the studied Na nanosystems, the parameters given in eqn. (5) and (6) were extracted from physical quantities recovered from the literature or numerically tested on single Na nanoparticles (see the ESI† for more details). The parameters finally exploited are the following: $\tau = 3.2 \times 10^{-14}$ s,[55] $\sigma_0 = 2.4 \times 10^7$ S m$^{-1}$,[56] $A_{ij} = 3.38$ Å$^2$, $l_{ij}^0 = 3.66$ Å,[56] $d = 12.00$, $s = 1.10$.

Before presenting the results and applications of ωFQ, some aspects of the newly developed approach need to be clarified. The equations specifying our approach are defined in the quasistatic regime, i.e. when the dimension of the studied system is much smaller than the wavelength of the external radiation. However, since the equations defining the model (see eqn (3)) can be rewritten in terms of Maxwell's sources (such as the electric current densities), its extension to the fully electrodynamical regime could be considered. This will be addressed in future communications and will permit the application of ωFQ to the calculation of the plasmonic response of nanostructures with a dimension comparable to the external radiation wavelength.

As stated before, here we focus on complex nanojunctions made of Na particles. Nevertheless, the model can be used in its present form for any metal for frequencies far from interband transitions. To back up this point, in section S4.2 in the ESI,† ωFQ is applied to selected silver nanorods of different sizes, for which the absorption maxima are far from interband transitions.[42] Almost perfect agreement with the reference ab initio data[42] is obtained. The inclusion of the effect of inter-band transitions is underway, in particular we are exploring both the inclusion of a Lorentz term in the conductivity and the explicit introduction of d-electron polarizability via an induced point dipole for each atom.

Furthermore, ωFQ has the potentiality to be coupled to electrodynamics models based on the metal permittivity, such as the Boundary Element Method (BEM), because the way the elements interact in both approaches is the same, i.e., via electromagnetic fields (see eqn (S11) given as the ESI†). In particular, this can be done by following the same approach which has been used to couple a polarizable MM layer to the Polarizable Continuum Model in the case of molecular systems.[57,58] The coupling of ωFQ with electrodynamical models will permit one to study much bigger systems, for instance by treating the core of nanoparticles with a continuum approach and retaining the atomistic description only for their surface.

## 3 Results and discussion

In order to test our newly developed ωFQ method, based on the Drude model for conductivity in metals, classical electrodynamics and quantum tunneling, we shall compute the optical response of Na aggregates which are characterized by subnanometer gaps. We shall compare the results obtained by exploiting our model with those calculated at the ab initio level.

In particular, the optical absorption spectra of a metal nanorod pulled beyond the breaking point[46] and those of two small metal nanoparticles brought into contact[47] are studied because they are paradigmatic of a class of nanoplasmonic problems where ab initio simulations seem mandatory.

As stated before, ωFQ equations are solved for each frequency given as the input. In the two considered cases, the absorption cross sections for all structures were calculated by considering 300 frequencies (from 0.0 to 3.0 eV, step 0.01 eV) and 450 frequencies (from 0.0 to 4.5 eV, step 0.01 eV) for the stretched nanorod and the nanoparticle dimer brought into contact, respectively. The total computational cost for each considered structure of the two systems is 11 seconds (11 Mb RAM) and 22 minutes and 55 seconds (112 Mb RAM), respectively. The calculations were performed on a MacBook Pro 2011, 2.3 GHz Intel Core i5, 4GB RAM. 4 cores were used for OpenMP parallelization. While we have no data on the time and memory requirements of the original ab initio calculations, supercomputers are certainly in order.

This is of course an important practical advantage of this approach compared to ab initio. To further deepen this point, in section S4 of the ESI† we compare the ab initio vs. ωFQ requirements for a Na$_{59}$ nanoparticle (see Table S3†). There we also show that the developed approach, although not fully algorithmically optimized, is already able to treat a 10 nm nanoparticle (around 13 803 atoms) with reasonable computational effort.

### 3.1 Stretched sodium nanorods

In this section the ωFQ approach (see Section 2) is applied to a challenging system, *i.e.* a mechanically stretched sodium nanorod, which has been recently studied at the *ab initio* level.[46] For such a system, absorption cross sections as a function of the elongation distances at the full *ab initio* level have been reported,[46] and such data are taken in this paper as reference values to evaluate the quality of our fully atomistic, but classical ωFQ approach. It is worth noting that upon increasing the elongation distance, a sub-nanometer junction region occurs, in which quantum tunneling effects play a crucial role in determining the spectral features.[32,41,43] Therefore, the application of our model to such a challenging system will highlight its potentialities and limitations at describing such effects.

The nanorod structures, eight of which are depicted in Fig. 1, were kindly provided to us by the authors of ref. 46. A total of 60 different structures were obtained from an initially perfect $Na_{261}$ nanorod, which was adiabatically stretched, by allowing the atomic positions of the central region to relax (see ref. 46 for more details).

For all 60 structures, absorption cross sections were calculated by exploiting the ωFQ model; Fig. 2 reports the absorption spectra of selected 30 structures as a function of the elongation distance.

As depicted in Fig. 1, the sodium nanorod is elongated and the atoms in the nanojunction region are relaxed until the structure breaks for distances longer than 26 Å, where the limit of mono-atomic junctions is reached.

These structural features are reflected by the calculated spectra (see Fig. 2(a)); in fact, a clear discontinuity is evident at $d = 26$ Å (structure G). Let us focus on elongation distances $d < 26$ Å. The pristine nanorod structure (structure A) presents one intense excitation at 1.5 eV (dubbed Local Plasmon LP) and a less intense peak at 2.8 eV (LP2). Our atomistic model allows the identification of the nature of such LPs, for instance by graphically plotting the imaginary part of atomic charges for each transition. Maps of the molecular electrostatic potential (MEP) obtained from such charges are reported in panel (a) of Fig. 3 for structures A–H. The comparison of data in panels (a) of Fig. 2 and 3 clearly shows a first charge-transfer excitation and a second transition with a dipolar character. Therefore, by exploiting the same nomenclature used for metal dimers, LP will be renamed as a Charge Transfer Plasmon (CTP), whereas LP2 as a Boundary Dipolar Plasmon (BDP).[59–61] As the elongation distance increases, both CTP and BDP significantly redshift, and this feature is particularly evident for CTP. In addition, they behave in a completely different way: CTP intensity slowly decreases, whereas BDP becomes more and more predominant. Such a behavior is commonly identified in most nanoplasmonic dimers.[45–47,62,63]

When the elongation distance reaches 26 Å (structure G) a monoatomic junction is obtained, which is the limiting structure occurring just before the structure breaks. Such features are reflected by the absorption cross section. The nature of the lowest energy transition is different from the previous CTP because the node is not placed at the geometrical center of the nanostructure, although it preserves the CT character (see Fig. 3, panel a). For this reason, such a plasmon is called CTP2. CTP2 occurs at about 0.5 eV and shows a very low intensity, because electrons can only transfer through a single atom. The BDP excitation becomes the most intense and shifts at 2.2 eV. The inspection of panel (b), structure G in Fig. 3 shows that such an excitation has now an octupolar character, characterized by 3 nodes. At such an elongation distance, a third excitation, which is actually already visible at 25 Å, arises at about 1.4 eV. The analysis of the MEP map suggests this transition to be due to an additional dipolar plasmon, BDP2.

We move now to comment on the spectra for $d > 26$ Å. CTP disappears, as expected, because the gap between the two nano-moieties is too large to allow electron tunneling between them. Thus, only BDP and BDP2 excitations are present. In particular, BDP2, which has a clear octupolar character as evidenced by the pictures reported in panel (b) of Fig. 3 for structure H ($d = 28$ Å), increases in intensity, although BDP still dominates the spectrum.

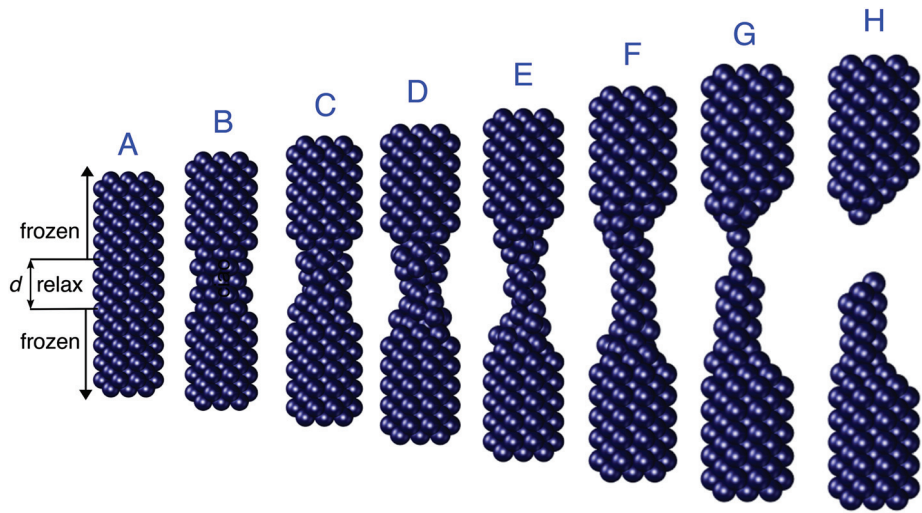

**Fig. 1** Selected structures obtained by Rossi *et al.*[46] by stretching a $Na_{261}$ nanorod. The elongation distances $d$ of the depicted structures are (from A to H): 0, 6, 10, 14, 20, 22, 26, and 28 Å.

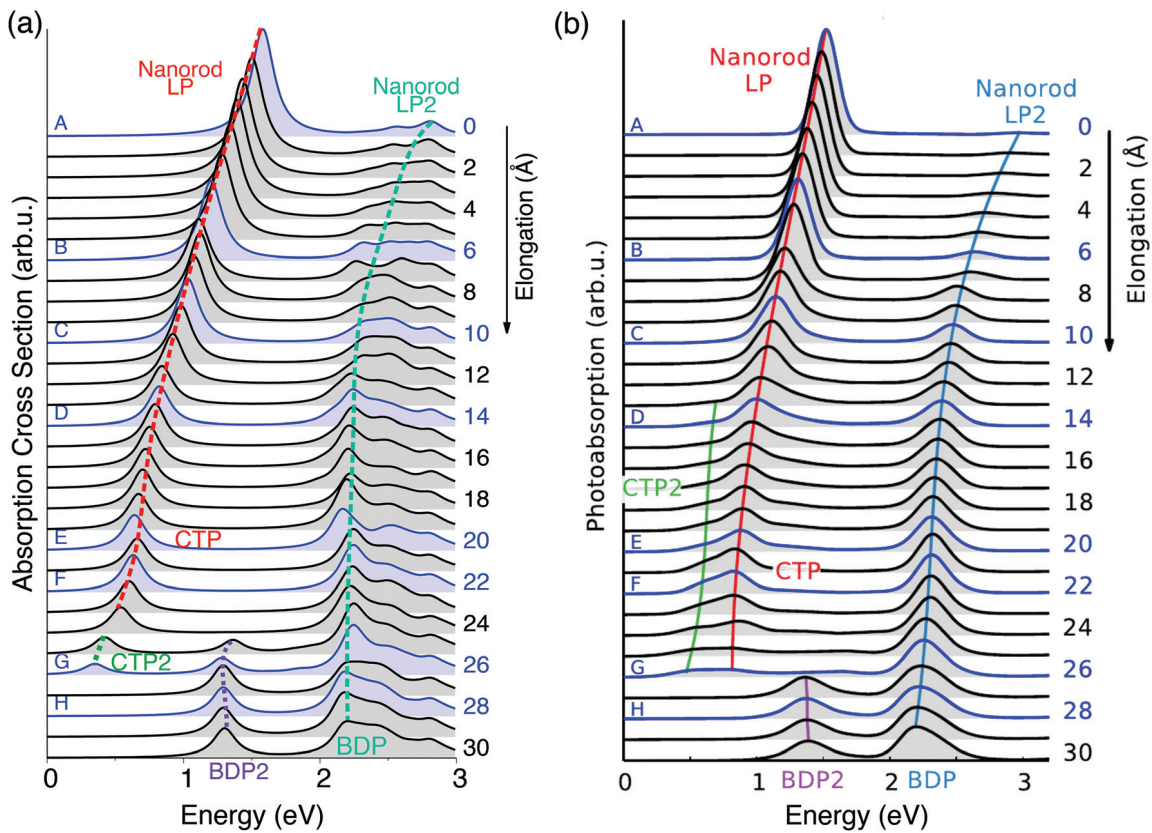

**Fig. 2** Evolution of the plasmonic response of a Na$_{261}$ nanorod under stretching. Structures A−H in Fig. 1 are highlighted. (a) ωFQ absorption cross section as a function of the energy. (b) Adapted with permission from ref. 46.

To better show the capabilities of our atomistic classical model, our results can be directly compared with the theoretical *ab initio* calculations reported in ref. 46, where most of the physical findings were fully disclosed. The resulting spectra are reproduced in panel (b) of Fig. 2. Our calculated absorption cross sections (panel (a) in Fig. 2) compare extremely well with the *ab initio* data, and only minor discrepancies are present. The *ab initio* spectra for structures with $13 < d < 26$ Å show a really low peak at about 0.5 eV, associated with CTP2. Such a peak, which however can hardly be identified in the *ab initio* density maps (see panel (b) of Fig. 3), is reproduced by ωFQ only in the case of the structures with $24 < d < 26$. Thus, ωFQ is not able to reproduce the relative intensity between CTP and CTP2, which are both present in the *ab initio* results. In addition, the *ab initio* BDP transition results in a narrower band. Such a difference can be justified by the atomistic nature of our model, which results in some kind of inhomogeneous broadening due to transitions with different nodal structures at the atomic scale, but corresponding to the plasmons of similar nature. Note that such inhomogeneous band broadening is not reported in the *ab initio* data probably because of the real-time TDDFT broadening. Furthermore, with reference to structure G, ωFQ predicts BDP to have a multi-polar character (with 5 nodes), whereas its *ab initio* counter-part shows such character only at higher energies (see Fig. 3, panels a and b). Such a discrepancy shows that for structure G, ωFQ tends to overestimate the 5 node plasmon character. However, it is worth pointing out that such a qualitatively different description is reported only in the case of structure G. Despite such minor discrepancies, the agreement between ωFQ and DFT spectra is impressive. In fact not only excitation energies but also relative intensities are correctly reproduced in all the elongation ranges, as well as the discontinuities that are reported for both CTP and BDP. Furthermore, ωFQ reproduces the *ab initio* calculated discontinuity (redshift) in the spectrum of the structure with $d = 7$ Å. Such a behavior is probably due to the structural rearrangement of the nanorod as a result of the *ab initio* geometry relaxation. The peculiar atomistic nature of ωFQ makes it capable of also exerting such effects, resulting from tiny deformations of the nanostructure.

To further analyze our results, in Fig. 4, excitation energies and integrated intensities calculated by exploiting our model are shown for the three plasmons. Many discontinuity points are noticed for both CTP and BDP as a result of the stretching

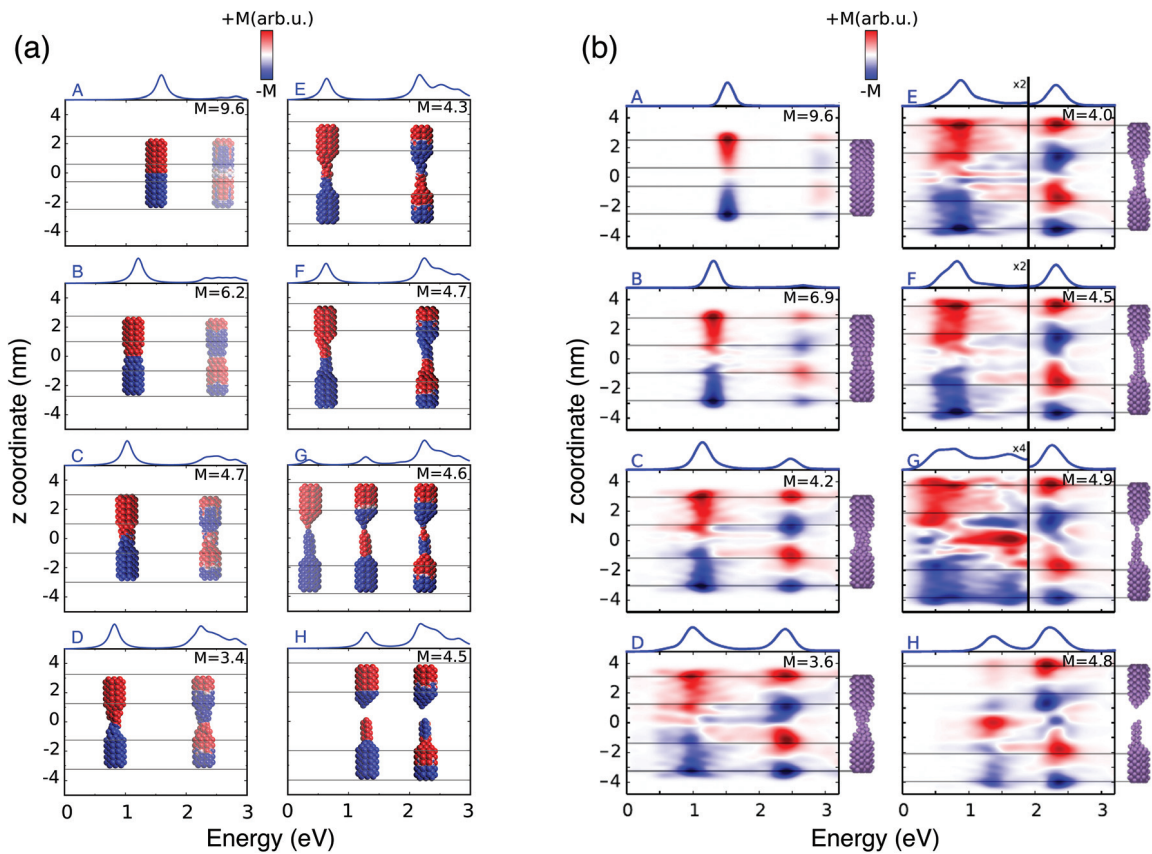

**Fig. 3** (a) ωFQ pictorial representation of the local plasmonic response for the 8 A–H selected structures. The blue color indicates a negative charge, whereas the red color indicates a positive charge. (b) Real time TD-DFT results adapted with permission from ref. 46.

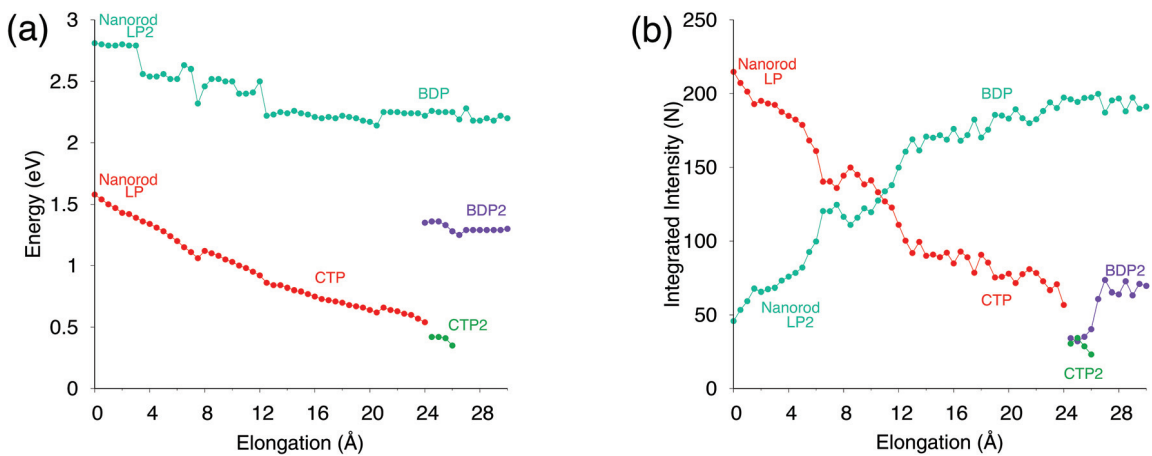

**Fig. 4** Analysis of the plasmon resonances during the nanorod elongation. (a) Energy maxima and (b) integrated intensity (*i.e.*, the area underlying the curve) of the plasmon modes as a function of *d*. The intensities are normalized so that the full spectrum integrates to the number of valence electrons (261).

of the nanostructure. In particular, at about 7.5 Å the energies of CTP and BDP decrease by 0.1 eV. Integrated intensities also present a discontinuity point at such a distance. Such shifts and discontinuities, which as stated before are due to struc-

tural rearrangements, are also reported by DFT calculations (see Fig. S6 given as the ESI†).[46]

Our results, which also in this case are quantitatively comparable to DFT, show once again that our classical atomistic

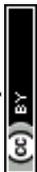

approach gives a correct description of the underlying physical phenomena.

### 3.2 Sodium NP dimer: approaching and retracting processes

As a second test to analyze the performances of ωFQ, the latter is challenged with the description of the optical properties of two $Na_{380}$ icosahedral nanoparticles which are approached and retracted (see Fig. 5). This system has been recently studied at the full *ab initio* level by Marchesin *et al.*,[47] who kindly provided us with the full set of model structures.

Two alternative processes will be considered: first, the two $Na_{380}$ nanoparticles are placed at a distance of 16 Å (such a distance guarantees that they do not interact) and then they are drawn closer until they fuse (see Fig. 5, panel a). Then the two fused $Na_{380}$ nanoparticles are retracted until the structure separates (see Fig. 5, panel b), giving rise to a process which is similar to the case presented in the previous section. Two alternative situations of approaching and retracting were both considered, because as has been already reported in ref. 47 the two processes are physically different.

Let us start the discussion by considering the approaching process (Fig. 5, panel a). The imaginary part of the longitudinal polarizability, *i.e.* the component parallel to the dimer axis, has been computed as a function of the inter-nanoparticle distance; its values are reported in panel (a) of Fig. 6.

The calculated 2D plots present a clear discontinuity between the nominal gap sizes of 6.1 Å and 6.2 Å, *i.e.* between structures B and C in Fig. 5 panel a, which correspond to a jump-to-contact instability. At higher inter-nanoparticle distances, plots are dominated by a single peak, which is placed at 3.19 eV at $d = 16$ Å, *i.e.* when the two nanoparticles are far apart. This band can be attributed to BDP. The induced charges and the corresponding MEP maps are depicted for the four selected significant structures in Fig. 7. We clearly see that for structure A BDP is a dipolar plasmon. As expected, as the distance between the two nanoparticles decreases, the BDP redshifts due to the increasing of electrostatic interactions. When the two nanoparticles fuse (structure C) a clear discontinuity appears and for $d \leq 6.1$ Å, the 2D plot is characterized by two main peaks, namely CTP (1.66 eV) and CTP′ (3.15 eV), at $d = 6.1$ Å, corresponding to the plasmon excitation represented for structures C and D in Fig. 7. As the distance further reduces, CTP blueshifts whereas CTP′ remains almost unchanged.

The inspection of the MEP maps in Fig. 7 shows that the higher order CTP′ shows a dipolar character, similarly to BDP, which occurs for structures with $d > 6.1$ Å. The jump-to-contact

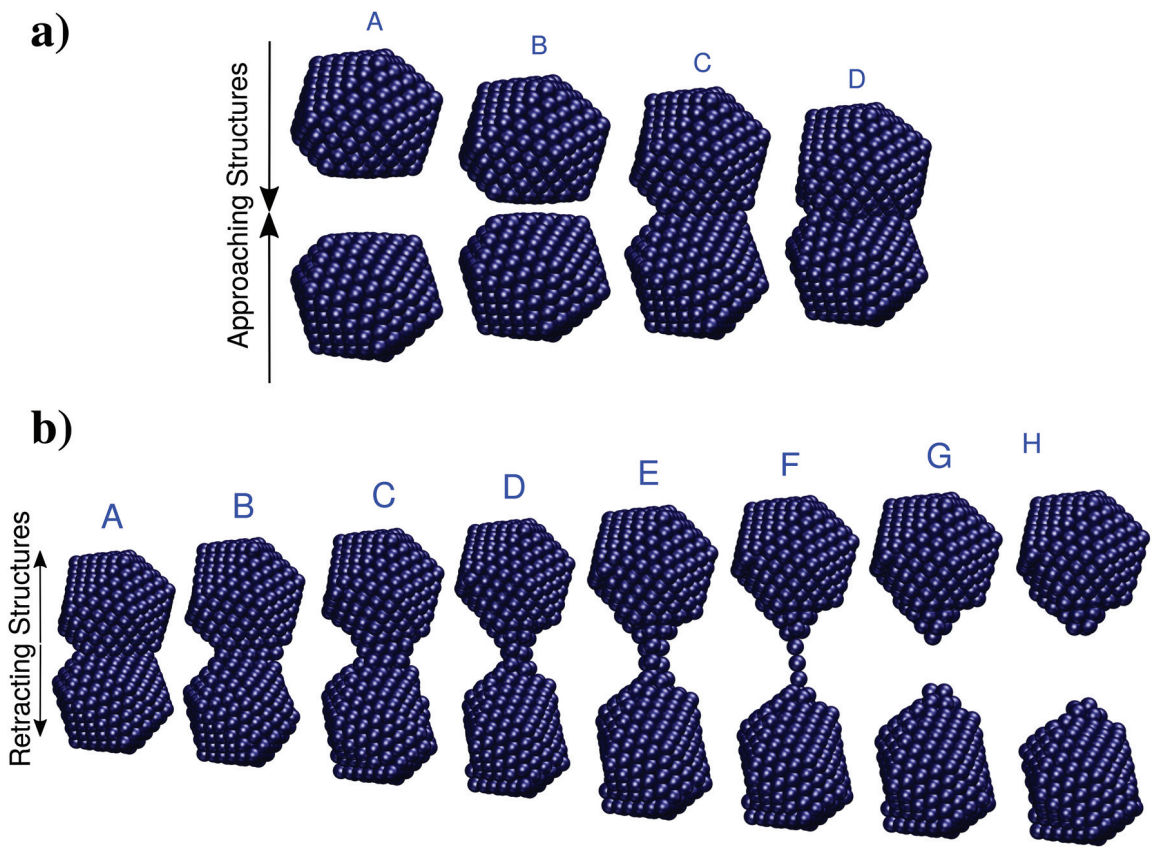

**Fig. 5** (a) Selected structures resulting from the approaching of two $Na_{380}$ nanoparticles. A: $d = 16$ Å; B: $d = 6.2$ Å; C: $d = 6.1$ Å; A: $d = 0$ Å. (b) Selected structures resulting from the retracting of two $Na_{380}$ nanoparticles. The nominal gap distances $d$ are (from A to H): 3.7, 9.7, 14.7, 22.7, 25.9, 32.1, 32.3, 34.1 Å. Data are taken from ref. 47.

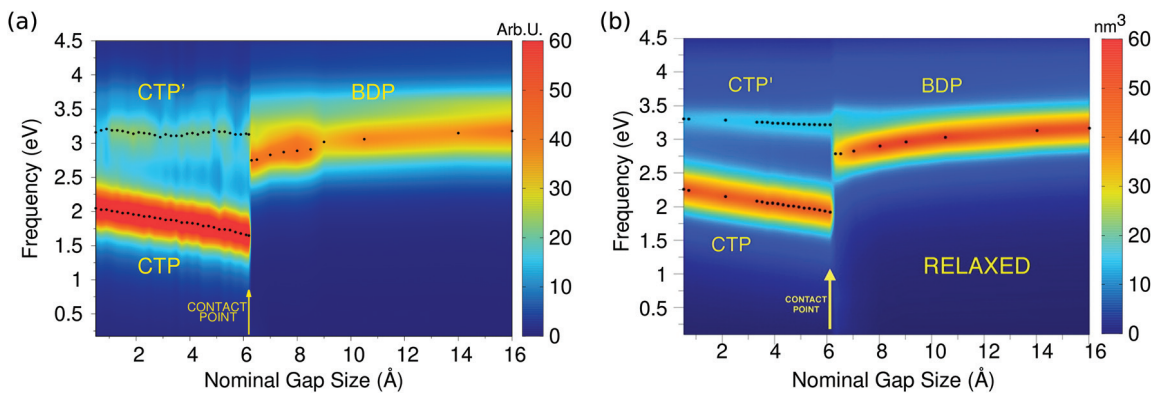

**Fig. 6** (a) ωFQ calculated 2D plots of the longitudinal imaginary polarizability (arbitrary units) as a function of the excitation energy and nominal gap size for the two approaching Na$_{380}$ nanoparticles. (b) *Ab initio* values adapted with permission from *ACS Photonics*, 2016, **3**, 269–277. Copyright 2016 American Chemical Society.

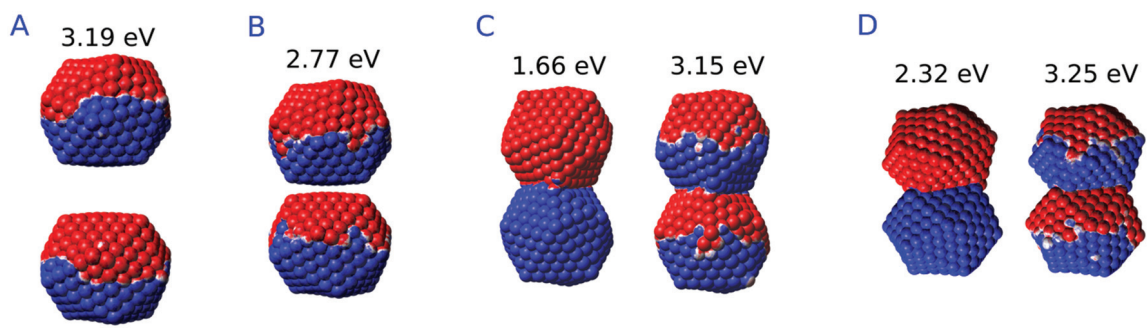

**Fig. 7** ωFQ MEP maps for plasmon excitation (eV) of selected structures. A: *d* = 16 Å; B: *d* = 6.2 Å; C: *d* = 6.1 Å; D: *d* = 0 Å. The blue color indicates a negative charge, whereas the red color indicates a positive charge.

structural instability is confirmed by the appearance of CTP, which is characterized by a net flux of charge between the two (fused) nanoparticles. Such a flux gives rise to a conductive regime, resulting in an electric current. Note that, as previously reported by Marchesin *et al.*,[47] the sudden occurrence of the junction bypasses the distance regime where quantum tunneling effects are relevant.

Moving back to Fig. 6, we note the very good agreement between the results obtained by exploiting the ωFQ approach and the *ab initio* counterparts. Qualitatively, DFT results are perfectly reproduced, in fact all the three CTP, CTP′ and BDP bands are described, their behavior as a function of the distance is correctly reproduced and the relative intensities of the bands are qualitatively well described. Some minor discrepancies are present from the quantitative point of view. In fact, the behavior of BDP as a function of the distance is not perfectly described, *e.g.* ωFQ intensities remain almost constant along the approaching process. Also, CTP intensities are overestimated and the CTP′ band seems broader. Such findings are in line with what has been found in the previous section and can be due to the atomistic nature of our approach, which does not smooth out inhomogeneities on the atomic scale.

We move now to study the two fused Na$_{380}$ nanoparticles which are retracted until the structure breaks (see Fig. 5 panel b for representative structures). As is evident the breaking process gradually occurs. In fact, a monoatomic junction arises (structure F), which breaks as the distance increases further. Therefore, tunneling effects are expected to be relevant, thus resulting in a different behavior of the calculated spectrum with respect to what we have reported in the previous paragraphs for the approaching process, and also found at the *ab initio* level.[47]

Indeed, this is confirmed by ωFQ calculated values of the imaginary part of the longitudinal polarizability, *i.e.* the component parallel to the dimer axis; such data are reported in panel (a) of Fig. 8 as a function of the elongation distance.

By starting from the fused A structure, we notice that, as expected, the spectrum consists of two bands, which can be related to CTP and CTP′ excitation. Their nature can be understood by referring to Fig. 9; CTP occurs at 1.84 eV and corresponds to a charge flux between the two nano-moieties. CTP′ (3.18 eV) shows instead the anticipated dipolar character by looking at the single nanoparticles.

As the elongation distance increases, both CTP and CTP′ redshift, and this is particularly evident especially for CTP. In addition the CTP band shrinks and its intensity decreases, whereas CTP′ shows an opposite behavior, *i.e.* its intensity increases and the band broadens. Small discontinuities, characterized by sudden red- or blue-shifts of the excitation,

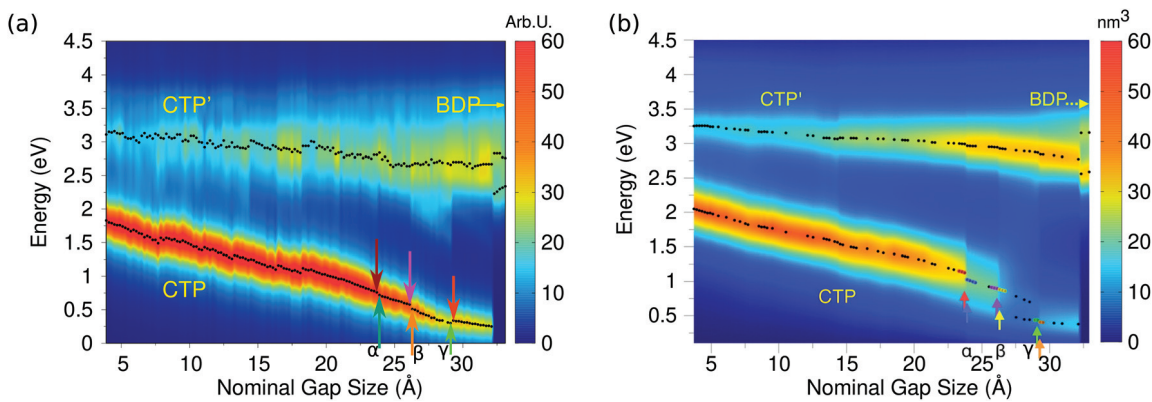

**Fig. 8** (a) ωFQ calculated 2D plots of the longitudinal imaginary polarizability (arbitrary units) as a function of the excitation energy and nominal gap size for the two approaching $Na_{380}$ nanoparticles. (b) *Ab initio* values adapted with permission from *ACS Photonics*, 2016, **3**, 269–277. Copyright 2016 American Chemical Society.

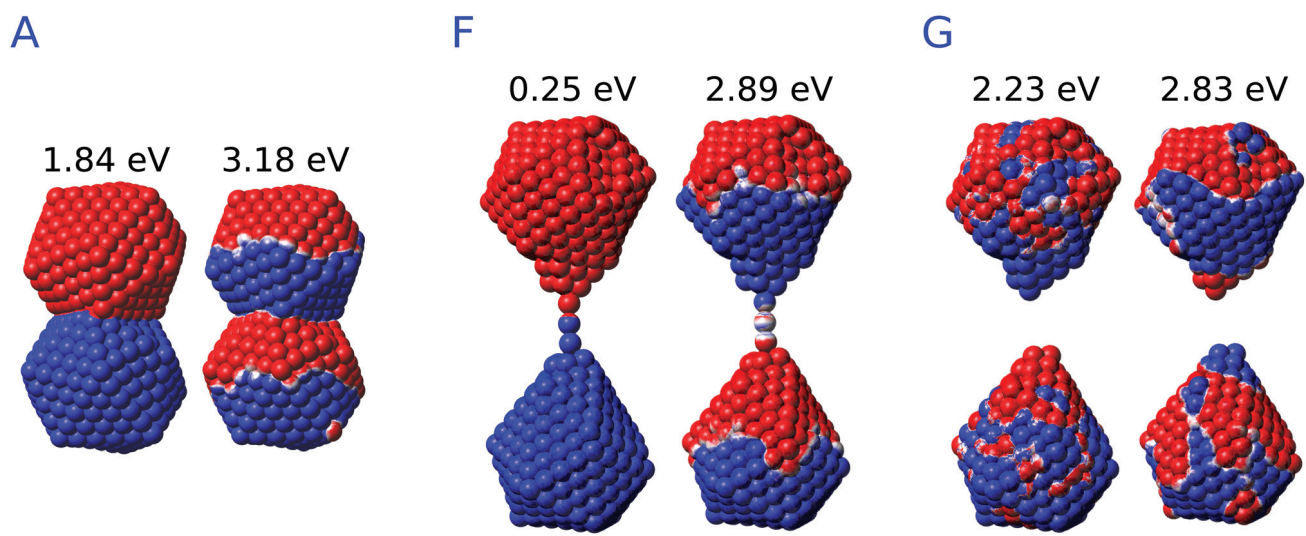

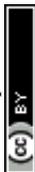

**Fig. 9** ωFQ MEP maps for plasmon excitation (eV) of selected structures. A: $d = 3.7$ Å; F: $d = 32.1$ Å; G: $d = 32.3$ Å. The blue color indicates a negative charge, whereas the red color indicates a positive charge.

are visible. This behavior is similar to what we have found in the previous section for the stretched $Na_{261}$ nanorod (see Fig. 4, panel (a)), and can be reasonably due to the structural relaxation and the resulting thinning of the conductive channels as the structure stretches. As the limiting structure F is reached ($d = 32.1$ Å), a monoatomic junction arises (see Fig. 9), resulting in the CTP band to occur at 0.25 eV and the CTP′ band at 2.89 eV. The inspection of the corresponding MEP maps shows that the nature of the associated plasmons is unchanged with respect to the initial A structure. Suddenly, the structure breaks (structure G, $d = 32.3$ Å), thus resulting in the disappearance of CTP and the convergence of CTP′ towards BDP. The MEP associated, depicted in Fig. 9, shows a multipolar character.

Moving back to Fig. 8, also for the elongation process a very good agreement between the results obtained by exploiting our classical atomistic ωFQ approach and the reference *ab initio* data[47] is noted. Qualitatively, DFT results are perfectly repro-duced, in fact all the three CTP, CTP′ and BDP bands are described, their behavior as a function of the distance is correctly reproduced and the relative intensities of the bands are qualitatively well described. ωFQ intensities for the CTP band are slightly overestimated, and remain higher also as the nominal gap size increases. Furthermore, ωFQ well reproduces the discontinuities in the spectra, and specifically those marked as α, β and γ in Fig. 8 panel (b). As already pointed out in the previous section, such a behavior can be due to the structural rearrangement of the nanostructure as a result of the *ab initio* geometry relaxation. The classical but atomistic nature of our approach makes it capable of correctly describing such effects.

The ωFQ imaginary charges for structures before and after the spectral jumps α, β and γ are depicted in Fig. S8, given as the ESI.† The structural change associated with each spectral jump is reflected by differences in the corresponding plasmons, *i.e.* by changes on the charges of the junction atoms.

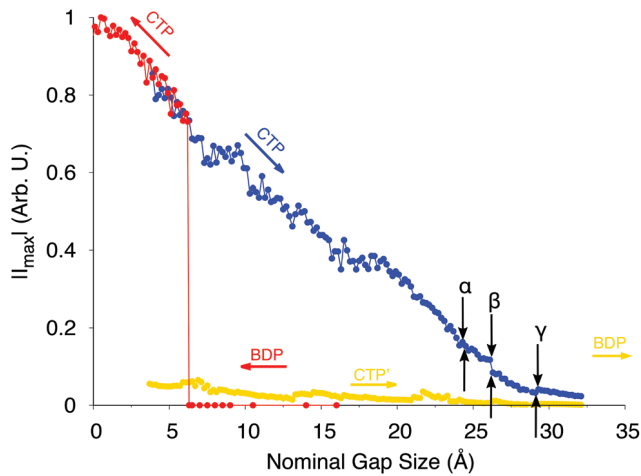

**Fig. 10** ωFQ absolute values of the electric current through the plasmonic nanojunction as a function of the elongation distance. Colored arrows indicate the direction of the process (approaching, in red, and retracting, in blue and orange). The current is computed by following ref. 47 (see also Fig. S9 given as the ESI†). Black arrows indicate the position of the α, β and γ spectral jumps. All values are given in arbitrary units.

Remarkably, our data are in agreement with DFT density distributions around the junction reported in ref. 47, thus showing once again the reliability of our classical atomistic model. Note again that most of the physical findings were fully disclosed in the reference *ab initio* study.[47]

To end the discussion and to further analyze the performance of the model, we report in Fig. 10 the calculated ωFQ absolute values of the electric current through the plasmonic nanojunctions as a function of the elongation distance. Both the approaching and retracting processes are considered. The reported values were obtained at the excitation energies of each plasmon.

As expected, for the approaching process when the two nanoparticles do not interact, i.e. when the spectra are dominated by BDP, no current flux is evidenced. As the jump-to-contact instability is reached, a discontinuity in the current arises, i.e. a net current flux is established. The current further increases as the inter-nanoparticle distance decreases.

For the retracting process, the CTP plasmon clearly dominates the charge flux. As the system is stretched, the current intensity slowly decreases, until it vanishes when the system breaks (structures F and G). Several discontinuities in the CTP current are present, similarly to what was already observed for the stretched nanorod in the previous section. Remarkably, the α, β and γ spectral jumps can easily be identified in the current plot. Note that the ωFQ plot reported in Fig. 10 can be directly compared to its *ab initio* counterpart depicted in Fig. S9 given as the ESI.†

## 4. Conclusions

In the present work, a novel atomistic model, ωFQ, based on textbook concepts (Drude theory, electrostatics, and quantum tunneling) has been proposed. In such a model, the atoms of complex nanostructures are endowed only with an electric charge, which can vary according to the external electric field. The electric conductivity between the nearest atoms is modeled by adopting the simplest possible assumption, i.e. the Drude model which has been reformulated in terms of electric charges. Thus, only a few physical parameters define our equations. Furthermore, the dielectric response of the system arises naturally from atom–atom conductivity. Remarkably, such a feature permits one to avoid the use of any experimental frequency-dependent dielectric constant, which is adopted in the quantum corrected models.[37] Moreover, ωFQ takes also into consideration quantum tunneling effects by switching off exponentially conductivity between neighboring atoms.

The ωFQ model was challenged to reproduce the optical response of complex Na nanoclusters which have been investigated previously at the *ab initio* level[46,47] and for which a QM description has been considered mandatory. The capability of our approach to reproduce the results of complex simulations has a relevant practical consequence; in fact, due to its classical formulation, ωFQ can be applied to model nanoplasmonic systems of size well beyond what can be currently treated at the *ab initio* level. Moreover, the good agreement between the *ab initio* simulations and ωFQ results shows that the physics it encompasses (Drude model, electrostatics and a quantum tunneling correction) properly ported at the atomistic level, is dominating the nanoplasmonic phenomena also in this small scale regime.

In this work, only Na clusters have been considered. However, ωFQ, properly extended to account for the atomic core polarizability that characterizes d-metals, has the potential to treat a great variety of plasmonic materials. Also, the formulation of the model in terms of electric charges and its manifest reliability shows that ωFQ has the potentialities to be coupled to fully QM molecular simulations within a QM/MM framework so as to allow the modeling of spectral enhancement of molecules adsorbed on plasmonic nanostructures. These aspects will be treated in future communications.

## Conflicts of interest

There are no conflicts to declare.

## Acknowledgements

We are greatly thankful to Prof. Risto Nieminen (Ålto University, Finland) and Prof. Daniel Sanchez-Portal (Centro de Fisica de Materiales CSIC-UPV/EHU Donostia-San Sebastian) for providing us the molecular structures of their studies. Financial support from SNS "Progetti Interni Coordinati 2016" is acknowledged. C. C. gratefully acknowledges the support of H2020-MSCA-ITN-2017 European Training Network "Computational Spectroscopy In Natural sciences and Engineering" (COSINE), grant number 765739. S. C. acknowl-edges funding from the ERC under the grant ERC-CoG-681285 TAME-Plasmons. We are thankful for the computer resources provided by the high performance computer facilities of the SMART Laboratory (http://smart.sns.it/).

# References

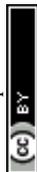

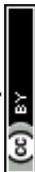